\documentclass[sigconf,nonacm,natbib=false]{acmart}

\AtBeginDocument{%
  \providecommand\BibTeX{{%
    \normalfont B\kern-0.5em{\scshape i\kern-0.25em b}\kern-0.8em\TeX}}}

\setcopyright{acmcopyright}
\copyrightyear{2022}
\acmYear{2022}
\acmDOI{10.1145/1122445.1122456}

\acmConference[DAC '22]{DAC '22: ACM Symposium on Neural
  Gaze Detection}{June 03--05, 2018}{Woodstock, NY}
\acmBooktitle{Woodstock '18: ACM Symposium on Neural Gaze Detection,
  June 03--05, 2018, Woodstock, NY}
\acmPrice{15.00}
\acmISBN{978-1-4503-XXXX-X/18/06}

\usepackage[style=ACM-Reference-Format,backend=bibtex,sorting=none]{biblatex}
\AtBeginBibliography{\footnotesize}
\usepackage{url}  
\usepackage{booktabs}  
\usepackage{multirow}
\usepackage{pifont}
\usepackage[caption=false,font=footnotesize]{subfig}
\usepackage[export]{adjustbox}

\usepackage{fancyvrb}

\newcommand{\pp}{\textsc{Prime+Probe}}

\newcommand{\fase}{\textsc{FaSe}}
\newcommand{\latctx}{\textsf{lat\_ctx}}
\newcommand{\asm}[1]{\texttt{#1}}
\newcommand{\reg}[1]{\textsf{#1}}

\definecolor{Yellow}{rgb}{1,1, 0.6}
\usepackage{colortbl}

\begin{document}

\pagestyle{plain}

\title{\textsc{FaSe}: Fast Selective Flushing to Mitigate Contention-based Cache Timing Attacks}

\author{Tuo Li}

\affiliation{%
  \institution{University of New South Wales}
  \city{Sydney}
  \country{Australia} 
}
 \email{tuoli@unsw.edu.au}
\author{Sri Parameswaran}
\affiliation{%
  \institution{University of New South Wales}
  \city{Sydney}
  \country{Australia}}
\email{sri.parameswaran@unsw.edu.au}

\thanks{This work has been submitted to the ACM for possible publication. Copyright
may be transferred without notice, after which this version may no longer be
accessible.}

\renewcommand{\shortauthors}{Li and Parameswaran, et al.}

\begin{abstract}

Caches are widely used to improve performance in modern processors. 
By carefully evicting cache lines and identifying cache hit/miss time, contention-based cache timing channel attacks can be orchestrated to leak information from the victim process. Existing hardware countermeasures explored cache partitioning and randomization, are either costly, not applicable for the L1 data cache, or are vulnerable to sophisticated attacks.
Countermeasures using cache flush exist but are slow since all cache lines have to be evacuated during a cache flush.
In this paper, we propose for the first time a hardware/software flush-based countermeasure, called fast selective flushing (\fase). By utilizing an ISA extension (one flush instruction) and cache modification (additional state bits and control logic), \fase\ \emph{selectively flushes cache lines} and provides a mitigation method with a similar effect to existing methods using naive flushing methods. \fase\ is implemented on RISC-V Rocket Core/Chip and evaluated on Xilinx FPGA running user programs and the Linux operating system. Our experimental results show that \fase\ reduces time overhead significantly by 36\% for user programs and 42\% for the operating system compared to the methods with naive flushing, with less than 1\% hardware overhead. Our security test shows \fase\ is capable of mitigating target cache timing attacks.


\end{abstract}




\maketitle

\section{Introduction}



\begin{figure}[tb]
\centering
\includegraphics[width=.72\columnwidth]{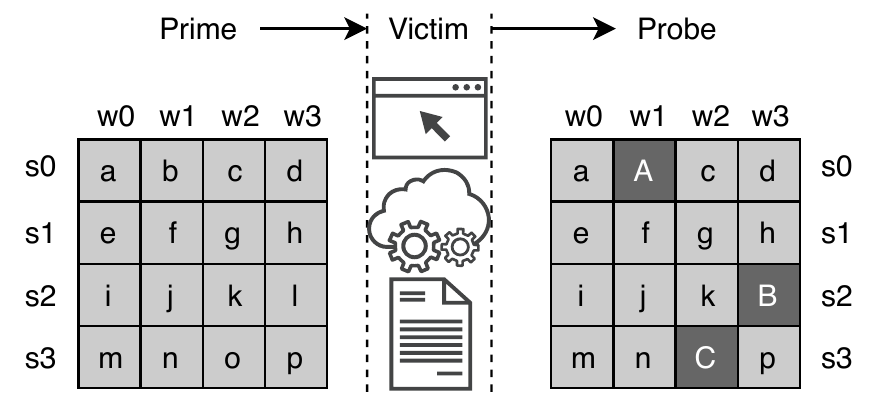}
\vspace{-10pt} 
\caption{Abstract view of a \pp{} attack. The blocks in dark grey represent the cache lines utilized by victim. \reg{si}: Cache Set i. \reg{wi}: Cache Way i.}
\label{fig:pp}
\vspace{-15pt}
\end{figure}

\begin{figure*}[tb]
\vspace{-5pt}
\centering
  \subfloat[b][Naive flush]{
\includegraphics[width=.20\textwidth]{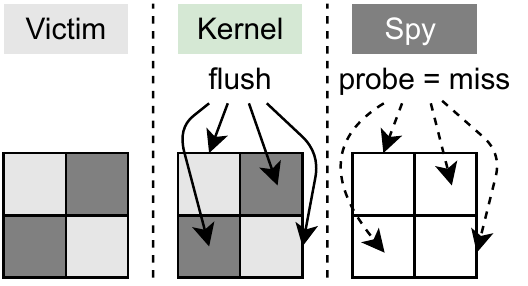} 
\label{fig:sysmodel}
  }
  \subfloat[b][Line level selective flush]{
\includegraphics[width=.20\textwidth]{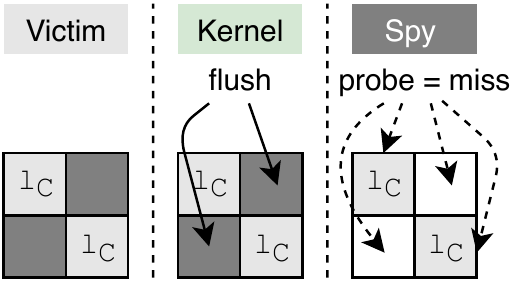} 
\label{fig:pcf}
  }
  \subfloat[b][Cache level selective flush]{
  \includegraphics[width=.19\textwidth]{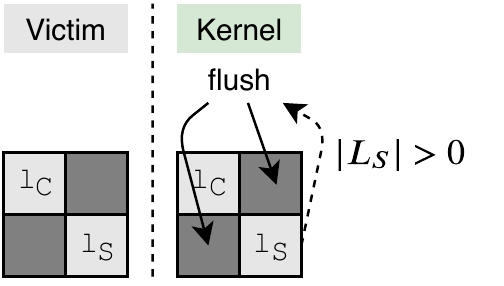} 
 \label{fig:fcf}
  }
  \subfloat[b][Flush cost estimation]{\footnotesize
  \begin{tabular}[b]{l c c}
  \toprule
  Flush Mechanism & Cost & Save\\
  \midrule
  Naive & 4 & n/a\\
  LLSF & 2 & 0.5 \\
  CLSF ($|L_S| \geq 1$) & 2 & 0.5\\
  CLSF ($|L_S| = 0$) & 0 & 1.0\\
  \bottomrule
  \end{tabular}
 \label{fig:cost}
  }
  \vspace{-8pt}
\caption{Brief illustration of \fase's key idea.}
\label{fig:idea}
\vspace{-10pt}
\end{figure*}

Cache timing side channel~\cite{Oren2015} has been a key component in recent lethal security attacks, such as Spectre/Meltdown,\footnote{\url{https://meltdownattack.com/}} on contemporary commodity processors.
Contention-based cache timing attack, e.g., \pp\ \cite{Osvik:2006:CAC}, is an important type of attack exploiting cache timing channels. In such an attack, an adversary utilizes carefully designed cache evictions, to learn the cache access pattern of the victim process. 

Fig.~\ref{fig:pp} demonstrates an example \pp\ attack, which has three stages. In \textbf{Stage 1} (prime), the spy process places its own data (\textsf{a} to \textsf{p}) in all the cache lines (in light grey). 
In \textbf{Stage 2}, the spy process gives up the processor core and waits for the victim process to execute its task on this core. During this stage, any cache usage on a cache line from the victim will result in contention at the corresponding cache line, and hence, will have attacker's prime data at these cache lines replaced and moved out of cache to the next level of cache-memory hierarchy. For example, in Fig.~\ref{fig:pp}, victim's data \reg{A}, \reg{B}, and \reg{C}, cause attacker's data \reg{b} (at \reg{s0,w1}), \reg{l} (at \reg{s2,w3}), and \reg{o} (\reg{s3,w2}) replaced.  
In \textbf{Stage 3} (probe), the spy process is switched back (simultaneously victim process is preempted from this core) to this core, and the spy proceeds to pinpoint the replaced cache lines, by loading or storing the prime data again. By measuring the time taken for accessing the prime data, the attacker can infer if this data access is a cache hit or miss, and hence, leak the information belonging to the victim process. The knowledge of such access patterns can leak critical information, e.g., secret key value in AES~\cite{Bernstein2005cache}, even across sandbox in web browsers~\cite{Oren2015}. 




L1 data cache is a critical microarchitecture in modern processors, which must be protected from cache timing attacks~\cite{Ge:2019:TPM}. Modifying cache architecture for partitioning~\cite{Domnitser:NMC:2012} and randomized cache mapping~\cite{Qureshi2018} is costly for local private caches and is vulnerable to sophisticated exploits~\cite{song2021randomized}.
State-of-the-art software-based mitigation methods~\cite{Zhang:2013:DRC,Oleksenko:2018:VPS,Ge:2019:TPM} mitigate timing attacks on private L1/L2 caches by performing \emph{cache flush} upon preemption, process switch, and syscalls.\footnote{\url{https://lwn.net/Articles/768418/}}
Cache flush is effective because it guarantees that the entire cache will be cleaned before the processor is switched from one process to another. After a cache flush, the hit/miss time difference can no longer be observed by the attacker's process, which enforces that the processes are isolated temporally.


While cache flush is effective in protecting against cache timing attacks, it significantly increases the cache miss rates and cold cache effects, key factors affecting program performance~\cite{Agarwal1989}.
Built upon cache flush, existing mitigation methods incur substantial program slowdown (more than 19\% throughput overhead in Nginx in~\cite{Oleksenko:2018:VPS}). 
Moreover, cache flush is likely to incur greater performance overheads in future processors. In fact, the performance cores (named Firestorm) in the recently released Apple M1 chip, contain  L1 data caches as large as 128 KB.\footnote{\url{https://en.wikipedia.org/wiki/Apple_M1}}


In this paper, we propose a novel hardware/software method called fast selective flushing (\fase), for countering contention-based cache timing attacks in L1 data cache. 
\fase\ collectively leverages a customized cache microarchitecture and a specialized cache flush instruction to create two new flush mechanisms, which significantly reduce the flush-based mitigation time cost. 
First, \emph{line level selective flush (LLSF)} mechanism is proposed, which allows flushing of a subset of the cache lines, instead of all cache lines in the cache, while guaranteeing that the cache hit/miss time difference cannot be observed by the attacker. 
Second, \emph{cache level selective flush (CLSF)} mechanism is proposed, which enables the cache flush to be guided (by user using a simple programming interface) and to perform ``strategically''  when necessary (when user-defined critical data is in cache), rather than always. 

\fase's hardware design extends the base processor architecture to support a specialized new cache flush instruction (\asm{scflush}), as well as extends the L1 data cache with minimum additional state bits and control logic to perform both the LLSF and CLSF mechanisms. The software support for \fase\ only takes a few lines of assembly code to: 1) integrate the specialized flush instruction in the software stack; and, 2) instrument program source code to define the critical data access. 
\fase\ is implemented and evaluated on a RISC-V ISA processor, called the Rocket Core with Rocket Chip system-on-chip (SoC), on a Xilinx ZYNQ Ultrascale+ FPGA. 
The microbenchmark and OS evaluation results show that \fase\ reduces 36\% (user programs) and 42\% (OS context switch latency) time overhead for flush-based mitigation on average. FPGA synthesis result shows that the hardware overhead of \fase\ is negligible (less than 1\% in tile with FPU excluded).

The contribution of this research is summarized as follows.
\begin{itemize}
\item To the best of our knowledge, \fase\ is the first selective flush-based mitigation for contention-based cache channel.
\item A RISC-V processor with \fase\ hardware and software modifications are implemented and validated on FPGA running user programs and Linux operating system.
\item A contention-based cache timing attack evaluation, using \pp, is performed on the \fase\ processor.
\end{itemize}


\section{Related Work}\label{sec:bgrw}

\textbf{Hardware-based} mitigation methods~\cite{Domnitser:NMC:2012,Yan:SHA:2017} modify cache architecture by partitioning cache lines among processes, to counter cache timing channels. 
These partitioning-based methods result in significant cache under-utilization, and hence, are not practical for local private caches, which are small and time-shared. Other hardware-based methods~\cite{Wang2008,Qureshi2018,Werner2019} explore using randomized memory-to-cache mapping in cache hardware. 
These randomization-based caches typically incur substantial hardware modifications, such as adding a mapping table (1/8 to 1x of cache size)~\cite{Wang2008,Qureshi2018}, index table~\cite{Werner2019}, and cryptography circuitry~\cite{Qureshi2018,Werner2019}. 
In addition, software modifications are required in OS, including grouping tasks~\cite{Qureshi2018}, extending page table~\cite{Werner2019}, and additional kernel/user communications for passing unique process identification (PID)~\cite{Werner2019}.
Randomized caches are prone to be breached if the attacker is provided with sufficient trials~\cite{song2021randomized}.
\textbf{Software-based} spatial partitioning~\cite{Kim2012,Liu:CDL:2016,Dong2018} are used to mitigate timing attacks on shared caches, such as Last-level caches (LLC).
For private L1/L2 caches, due to the high cost of cache partitioning~\cite{Ge:2019:TPM}, software-based methods~\cite{Zhang:2013:DRC,Oleksenko:2018:VPS,Ge:2019:TPM} are proposed to flush caches, when the processor core is about to run another user's thread or kernel thread. These methods typically require considerable software modification of the OS kernel (e.g., 1.4K LoC in~\cite{Zhang:2013:DRC}). 
\textbf{Compound flush instructions} are used in~\cite{bourgeat2019,Wistoff:PMC,Li:SIMF} to flush multiple on-core microarchitectural states, including L1 caches, TLB, branch prediction unit. This flush is a superset of naive cache flush, which flushes every cache line.
These flush instructions are built on RISC-V processor variants, whose cache architectures and coherence protocols are different from the processor (Rocket Chip) used in this paper. For example, in~\cite{Wistoff:PMC}, L1 data cache uses write-through policy, which does not require write-back during cache flush. Rocket Chip has a write-back L1 cache, which is more popular in modern processors. 
Compared to prior arts, \fase\ is the first selective flush-based method to mitigate cache timing attacks. Such selective flushing reduces overheads significantly.


\section{\fase\ Selective Flush}\label{sec:concept}

\textbf{Threat Model.}
In this paper, contention-based, on-core L1 data cache timing attacks are targeted, which have been investigated in prior arts~\cite{Qureshi2018,Zhang:2013:DRC,Oleksenko:2018:VPS,Ge:2019:TPM}.
The adversary is assumed to have user rights and mounts the spy process running on the same processor core. In this paper, \pp\ is used as the reference contention-based cache timing attack.
Other levels of cache hierarchy are out of scope and assumed to be protected from cache timing attacks using existing methods discussed in Section~\ref{sec:bgrw}.

\textbf{Nominal flush-based mitigation system model.}
A system with flush-based mitigation executes cache flush in between user processes at OS kernel~\cite{Ge:2019:TPM} or enclave exit~\cite{Oleksenko:2018:VPS} (i.e., flush point). 
Fig.~\ref{fig:idea} briefly illustrates existing and our proposals in comparison.
As shown in Fig.~\ref{fig:sysmodel}, in a flush-based system, the CPU events of one CPU core can be viewed as interleaved cache flush events and compute/service events.
Naive cache flush mechanism (Fig.~\ref{fig:sysmodel}) \emph{flushes all cache lines of the local private data caches}. Such a naive mechanism can either be implemented in software (e.g., run \asm{dc\textvisiblespace cisw} instruction in for loop in ARMv8 ISA) or hardware (e.g.,~\cite{bourgeat2019,Wistoff:PMC}), essentially in a for-loop, which traverses, invalidates, and cleans all cache lines in the cache. After a flush event, all the following memory accesses will be cache-miss and take a much longer time.

\textbf{Our proposal 1: not every cache line requires a flush.}
In contrast to the naive flushing, the first method proposed here (called \emph{Line Level Selective Flush} or \textbf{LLSF}), only flushes the cache lines that were not updated in the current time slice. The intuition behind this method is that the updated cache lines will not result in a hit with the spy process, and thereby reducing the number of lines being flushed (this significantly reduces the flushing time and the time overhead associated with flushing). Like naive flushing, this method mitigates cache timing attacks and covert channels.
As shown in Fig.~\ref{fig:pcf},
at one flush point, let the process executed before this flush point be noted as current process $\tau_{cur}$ (victim in Fig.~\ref{fig:pcf}), while the process executed after this flush point be noted as next process $\tau_{nxt}$ (spy in Fig.~\ref{fig:pcf}).
Let's denote the number of total cache lines as a set $L$ and the number of dirty cache lines as $L_D$ ($L_D \subseteq L$).
There is a subset of cache lines $L_C$ ($L_C \subseteq L_D$), which have been placed into the cache by $\tau_{cur}$. Flushing the complementary of these cache lines, noted as $L_X = L_D \setminus L_C$, is sufficient to let $\tau_{nxt}$ observe cache miss at every cache line $l \in L$. Therefore, $\tau_{cur}$ and $\tau_{nxt}$ are temporally isolated while fewer cache lines need to be flushed. A decrease in flushed cache lines leads to the reduction of flush cost and alleviates the cold cache penalty (elaborated later). 

\textbf{Our proposal 2: not every flush point needs a cache flush.}
In the second proposed method, \emph{cache level selective flush} (\textbf{CLSF}) the program section that requires protection is instrumented (for example with pragmas), and only if that part of the program brings data into the cache, then the cache is flushed. 
The intuition behind the second method is that there are many parts of the program which do not require temporal isolation for countering cache timing attacks, and experience shows a program executes in multiple OS time slices, and only some of these deal with protected data, such as an encryption key. This method mitigates the cache timing attacks, but will not fully counter cache covert channels, if the covert channel is implemented using the program parts that are not instrumented.
Fig.~\ref{fig:fcf} shows the brief idea of CLSF. Let us assume a program only has limited parts (e.g., T-table look-up in AES~\cite{Tromer:ECA:2010}), which are security-critical. We define the data accessed in security-critical parts of a program as security-critical data. At one flush point, if $\tau_{cur}$ (victim in Fig.~\ref{fig:pcf}) has not accessed any security-critical data in any cache lines (noted as $L_S$),
temporal isolation against $\tau_{nxt}$ at this point is unnecessary. Therefore, this cache flush can be nullified to trade flush cost with acceptable security degradation.


\textbf{Flush cost-saving analysis.}
The time cost of one cache flush event at time $t_i$ is mainly determined by the number of dirty cache lines $L_D$ ($L_D \subseteq L$) and the number of total cache lines $L$. This relation can be expressed as
$
P_i = \alpha \cdot |L_D^i|+\beta \cdot |L|
$
where the first term represents the time cost for invalidating and cleaning dirty cache lines, and the second represents the time taken to visit every cache line like a for-loop at flush.
$\alpha$ stands for the time penalty of invalidating and cleaning one dirty cache line, which is mainly the time taken for maintaining data coherency, such as write-back. $\beta$ denotes the time taken for traversing one cache line. $\alpha$ is order-of-magnitude larger than $\beta$. Flush cost is proportional to $|L_D|$.
Moreover, cache flush induces \emph{cold-cache penalty}, equivalent to the ``start-up effects''~\cite{Agarwal1989}. For a period of time, the total time cost of a flush-based system is 
$
\sum_i{P_i} + P_{CC}, i \in \{1, 2, \ldots , N\}
$
where $N$ is the number of total flush events and $P_{CC}$ is the lump sum of time penalties from cold cache effects.
Fig.~\ref{fig:cost} shows a brief quantitative comparison (based on flushed cache lines) between \fase\ and naive flush-based mitigation, corresponding to the scenarios in Fig.~\ref{fig:pcf} and Fig.~\ref{fig:fcf}. The flush cost is estimated as cache lines flushed. With LLSF, flush cost can be substantially reduced, if there are sizable $L_C$ cache lines that can avoid flushing. In this example, two out of four cache lines are saved from flushing, which leads to proportional amount of time cost saving. With CLSF, if the condition meets, i.e., $L_S = \emptyset$, the flush can be nullified and leads to a complete save of flush cost at this flush point.

\section{\fase\ Design: The Case for RISC-V}\label{sec:design}


\begin{figure}[tb]
\centering
\includegraphics[width=\columnwidth]{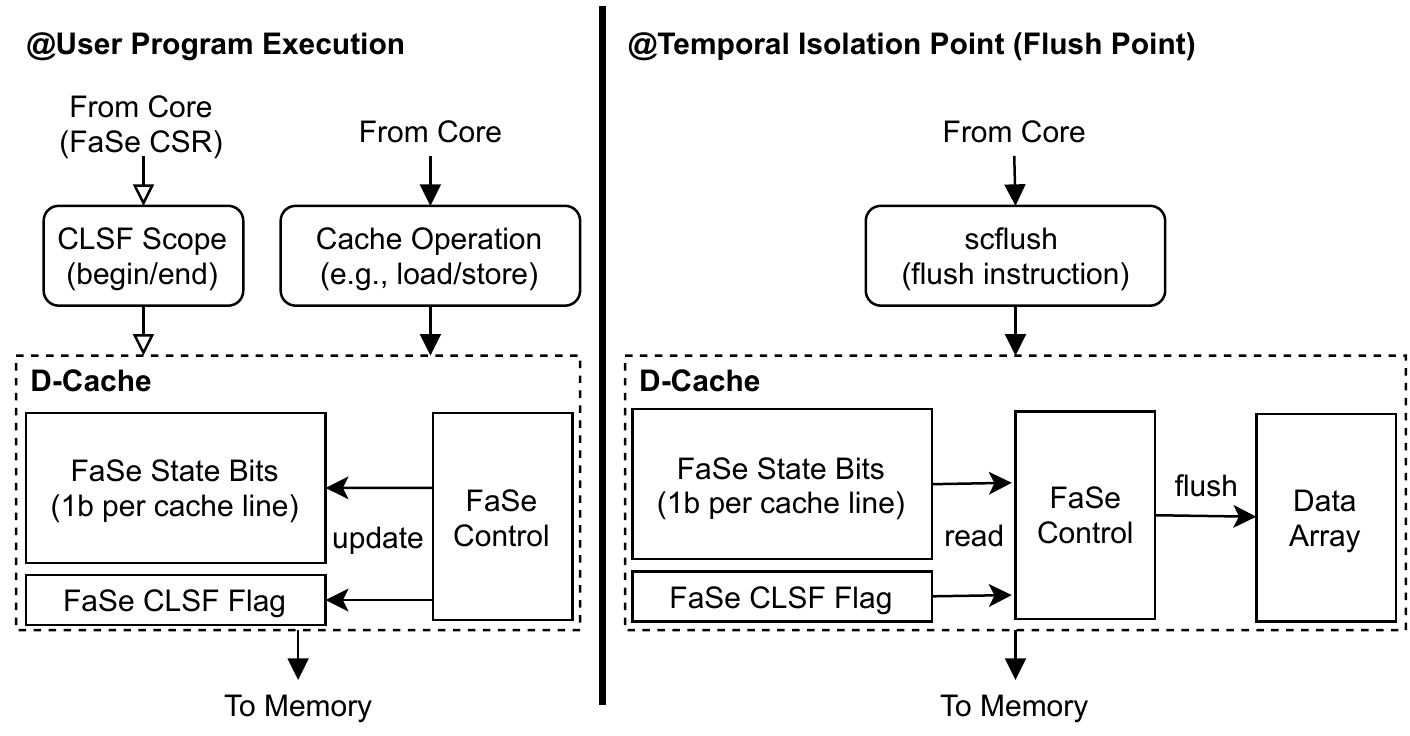} 
\vspace{-20pt}
\caption{\fase\ system overview.}
\label{fig:designOverview}
\vspace{-15pt}
\end{figure}

\begin{figure}[tb]
\centering
\includegraphics[width=0.75\columnwidth]{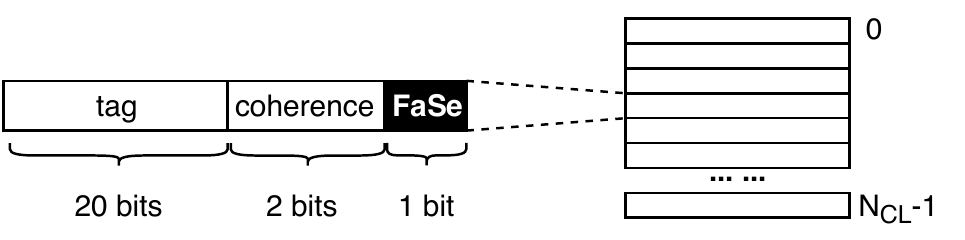} 
\vspace{-10pt}
\caption{\fase\ cache's tag array.}
\label{fig:tagArray}
\vspace{-15pt}
\end{figure}

Fig.~\ref{fig:designOverview} shows the overview of \fase\ design, which realizes both LLSF and CLSF. 
\fase\ design includes hardware modification to L1 D-cache, a cache flush instruction (\asm{scflush} - which is an ISA extension), and a control/status register (CSR), denoted as \reg{csr.scf}. \fase\ works in two phases, \emph{user program execution}, and \emph{flush instruction execution}.
\fase\ extends L1-D cache with 1) \fase\ state bits, 2) one CLSF flag status bit, and 3) \fase\ control.
Fig.~\ref{fig:tagArray} depicts \fase\ cache modification with respect to the  cache metadata (stored in tag array) in the RISC-V Rocket core processor. \fase\ adds one \fase\ state bit for each cache line. \fase\ state bits are stored together with other cache metadata, such as tag bits and coherence bits, which are one-on-one mapped to cache lines. In a 4-way 64-set data cache, the total number of \fase\ state bits is $1 \times 256$. Bit value ``1'' means this cache line has been accessed in the current process time slice (since the last cache flush). 

 \begin{figure}[t]
\centering
\subfloat[Flush algorithm]{
\includegraphics[width=.39\columnwidth]{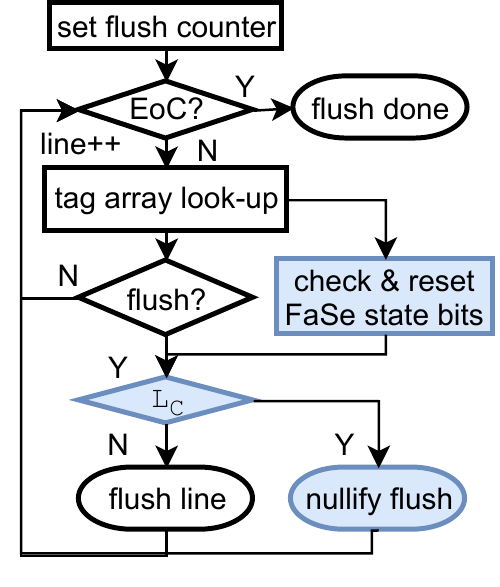} 
\label{fig:llsfFlow}
}
\subfloat[LLSF flush decision]{\footnotesize
  \begin{tabular}[b]{c c c}
  \toprule
  Coherence & \fase\ & Flush?\\
  \midrule
  \texttt{11 (M)} & \texttt{1} & N (nullify)\\
  \texttt{11 (M)} & \texttt{0} & Y\\
  \texttt{10 (E)} & \texttt{1} & N (nullify)\\
  \texttt{10 (E)} & \texttt{0} & Y\\
  \texttt{01 (S)} & \texttt{1} & N (nullify)\\
  \texttt{01 (S)} & \texttt{0} & Y\\
  \texttt{00 (I)} & \texttt{1} & N\\
  \texttt{00 (I)} & \texttt{0} & N\\
  \bottomrule
  \end{tabular}
  \label{fig:llsf_states}
}
\vspace{-8pt}
\caption{LLSF flush mechanism. EoC: end of cache lines.}\label{fig:llsfAlgo}
\vspace{-10pt}
\end{figure}

LLSF control mechanism has two parts, corresponding to \fase's two phases.
During user program execution, when CPU core accesses the  L1 cache, using instructions such as load or store, \fase\ state bit is updated to ``1''. This update is executed simultaneously when coherence bits are updated for the corresponding cache line by the native cache-control hardware.
Fig.~\ref{fig:llsfAlgo} illustrates LLSF algorithm when flush instruction is executed. Fig.~\ref{fig:llsfFlow} shows a flowchart of the steps in cache flush. Fig.~\ref{fig:llsf_states} depicts how LLSF decision is made based on coherence bits (assuming RISC-V's MESI-like coherence protocol) and \fase\ state bit of one cache line. In Fig.~\ref{fig:llsf_states}, Row 2, 4, and 6, are the cases where cache line flush is not necessary and hence avoided.
As shown in Fig.~\ref{fig:llsfFlow}, when the flushing instruction is executed, LLSF control mechanism sets the flush counter (from 1 to number of cache lines)   and performs the following steps for each cache line :
\textbf{\textcircled{\raisebox{-.8pt}1}} read \fase\ state bit and coherence bits of the current cache line from the tag array, using flush counter value as the tag array index;
\textbf{\textcircled{\raisebox{-.8pt}2}} check whether the coherence state of this cache line determines that this cache line should be flushed, and whether \fase\ state bit (shown in Fig.~\ref{fig:llsf_states} indicates this line flush is unnecessary;
\textbf{\textcircled{\raisebox{-.8pt}3}} reset \fase\ state bit of current cache line;
\textbf{\textcircled{\raisebox{-.8pt}4}} if \fase\ control determines that this cache line should be flushed (based on Fig~\ref{fig:llsf_states}), such as the situations in Row 3, 5, 7 in Fig.~\ref{fig:llsf_states}, cache line flush takes place, otherwise, this cache line flush is nullified; and,
\textbf{\textcircled{\raisebox{-.8pt}5}} increment flush counter, and if the counter value reaches maximum value (meaning last cache line has been processed), cache flush is finished.

\begin{SaveVerbatim}[commandchars=\\\{\}]{setfcf}
\textcolor{blue}{//CLSF critical segment begin}
\textcolor{blue}{asm volatile("csrwi scf, 1");}
set_key(key, key_len, enc, ctx);
\textcolor{blue}{//CLSF critical segment end}
\textcolor{blue}{asm volatile("csrwi scf, 0");}
\end{SaveVerbatim}

\begin{figure}[t]
\vspace{-10pt}
\centering

\subfloat[Code snippet]{\footnotesize
	\BUseVerbatim[fontsize=\footnotesize]{setfcf}
	\label{fig:setclsf}
}
\subfloat[Flush algorithm]{
\includegraphics[width=.5\columnwidth]{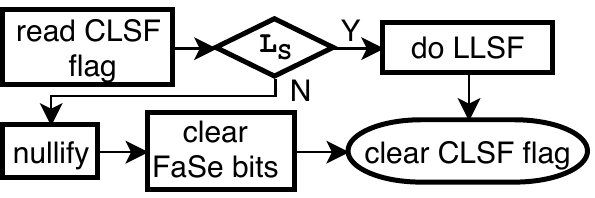} 
\label{fig:clsfFlow}
}
\vspace{-12pt}
\caption{CLSF flush mechanism.}\label{fig:clsfAlgo}
\vspace{-15pt}
\end{figure}


In addition to  LLSF, \fase\ CLSF further extends the CPU core with \fase\ CSR register (denoted as \reg{csr.scf}) and L1-D cache with one CLSF flag status bit.  \fase\ CSR is one-bit wide and programmable by the user. \fase\ \reg{csr.scf} allows users to mark the critical segment in the program code. When \reg{csr.scf}'s value is ``1'', the memory-related instructions during this time are considered critical and are protected.  When \reg{csr.scf}'s value is ``0'', instructions executed are no longer considered critical.
Fig.~\ref{fig:setclsf} depicts an example code snippet in AES program where CLSF critical segment is marked (assuming RISC-V ISA). In this code snippet, the user adds two additional lines of assembly code (inline assembly in C) to put the \reg{set\_key} function into CLSF critical segment. Before calling the \reg{set\_key} function, csr.scf is written with ``1'' by \asm{csrwi} instruction (In RISC-V ISA, one CSR register can be written using CSR access instructions, such as \asm{csrwi} and \asm{csrw} instructions). After returning from the \reg{set\_key} function, csr.scf is reset to ``0'' by \asm{csrwi} instruction. In L1 D-cache, CLSF flag status bit is added to indicate if any cache line has been used in the user-defined CLSF critical segment in current process time slice since last cache flush.

CLSF control mechanism has two parts with respect to user program execution and flush instruction execution.
During user program execution, CLSF looks at the signal value from \fase\ CSR  \reg{csr.scf}, when cache operation is issued from CPU core to L1 D-cache. If \reg{csr.scf}'s value is ``0'', the current cache operation is treated as non-critical. In this case, CLSF control does not do anything. If \reg{csr.scf}'s value is ``1'', the current cache operation is treated as critical. In this case, if any cache line's coherence state bits are updated due to this cache access, the CLSF flag status bit is asserted to ``1''.
Interruptions and exceptions could occur during a CLSF critical segment. To handle such a situation, when the CPU switches from the current process to kernel space, \reg{csr.scf} is saved as context and reset. When this process is switched back, \reg{csr.scf} is also restored as a part of the process context. This procedure incurs negligible additional code lines (less than 10 lines of assembly).
As shown in Fig.~\ref{fig:clsfFlow}, when \asm{scflush} is executed, CLSF control mechanism has the following steps:
\textbf{\textcircled{\raisebox{-.8pt}1}} read and check the CLSF flag value;
\textbf{\textcircled{\raisebox{-.8pt}2}} If CLSF flag value is ``1'', CLSF control nullifies this cache flush and clears \fase\ state bits. If CLSF flag value is ``0'', like LLSF, CLSF can examine \fase\ state bits and selectively flushes the cache lines, which are necessary; and,
\textbf{\textcircled{\raisebox{-.8pt}3}} At last, clear the CLSF flag.

\section{Experiment and Results}\label{sec:exp}




\textbf{Implementation.}
\fase\ is implemented on the RISC-V Rocket Core processor. Rocket Core is  part of the Rocket Chip SoC generator~\cite{Asan2016}. 
In our experiment, we use the 64-bit generic RISC-V ISA, \reg{RV64GC}.
The L1 data cache is a 32-KB 8-way set-associative cache containing 64-byte cache blocks.
We used an FPGA build\footnote{\url{https://github.com/ucb-bar/fpga-zynq}} of Rocket Chip and ported this build to Xilinx Ultrascale+ ZCU102 FPGA board. 
The main memory is a 4-GB DDR4 SODIMM manufactured by Micron fitted on to the ZCU102. 
The FPGA synthesis tool used for synthesis is Vivado 2017.1.

\textbf{Methodology.}
Our experiment mainly evaluates \fase's LLSF, given CLSF requires user and domain knowledge to set the protection scope. To showcase the efficacy of \fase's CLSF, we performed a case study on AES encryption. We compare our system against two reference systems in our experiments: 
1) the original \reg{RV64GC} Rocket Core, which is the \textbf{baseline} (without mitigation); and,
2) the \reg{RV64GC} Rocket Core augmented with hardware-supported naive cache flush (cache-flush with hardware for-loop similar to~\cite{bourgeat2019,Wistoff:PMC}),\footnote{Similar to CFLUSH.D.L1 \url{https://github.com/chipsalliance/rocket-chip/pull/1712}} which is denoted as the \textbf{naive} method or naive system.


In \textbf{Section~\ref{sec:krres}}, to observe how effectively \fase\ can reduce the OS overhead, we use context switch latency (one key overhead of flush-based method\footnote{\url{brendangregg.com/blog/2018-02-09/kpti-kaiser-meltdown-performance.html}}) test, \latctx{}, in LMBench 3.0~\cite{McVoy:1996:LPT} to evaluate overhead caused by \fase\ and naive systems. We modified the Linux kernel 4.20 by adding a cache flush instruction in the context switch routine for deploying \fase\ and the naive systems (on the software side). The baseline processor runs the native Linux kernel without modification. Naive and \fase\ systems run the modified Linux kernels. In this experiment, \fase\ enables LLSF. \latctx{} test runs with varied process sizes (by default from 0 to 64 KB). When process size is 0 KB, the process does nothing except pass the token on to the next process. Non-zero process size means that the process does some work (simulated as summing up an array) before passing on the token. For each process size, \latctx{} test considers a range of process numbers (by default from 2 to 96) and has context switch latency measured for each process number. Context switch latency is defined in LMbench as the time needed to save the state of one process and restore the state of another process. 
In \textbf{Section~\ref{sec:mbres}}, we evaluated the user program performance using programs from MiBench benchmark suite~\cite{Guthaus:2001:MFC}. In this evaluation, we use RISC-V proxy kernel (\reg{riscv-pk}).\footnote{\url{https://github.com/riscv/riscv-pk}} \reg{riscv-pk} is a basic application execution environment (basic POSIX syscalls and simple virtual memory management), which directly supports measuring cycle count and instruction count of the user programs. We modified \reg{riscv-pk} to adopt the \fase{}  method. In each program, temporal isolation takes place at user/kernel switch, which is for syscalls and exception/interruption handling.
In \textbf{Section~\ref{sec:secres}}, for security evaluation, we created a \pp\ cache timing attack based on the codebase from~\cite{Yarom:mastik}. We implemented the victim as a function in kernel space within \reg{riscv-pk}. The attacker process prepares the attack and uses a special \emph{syscall} to switch to the victim process. After the victim finishes the task in kernel space, the attacker process switches back and probes the cache. In comparison to mounting \pp\ attack on OS, this setup leads to a faster \pp\ attack, since the code executed between victim and spy is much less.



\begin{table}\footnotesize
\centering
\caption{Context switch latency (microseconds) across three systems, process sizes (SZ in kilobytes), process numbers (P$\cdot$)}\label{tab:latctx_full}
\vspace{-8pt}
\begin{tabular}{@{} c c r r r r r r r r@{}}
\toprule
System & SZ & P2 & P4 & P8 & P16 & P24 & P32 & P64 & P96 \\
\midrule 
\multirow{6}{*}[-0.5em]{Baseline} & 0 & 40.5 & 40.3 
	& 69.6 & 86.4 & 101.3 & 102.3 & 108.3 & 109.5\\
 & 4 & 51.0 & 88.0 & 119.6 & 139.3 & 138.0 & 142.8 & 144.2 & 144.4\\
 & 8 & 65.5 & 131.5 & 155.0 & 164.5 & 168.3 & 172.8 & 175.7 & 174.3\\
 & 16 & 140.5 & 202.3 & 213.3 & 225.8 & 228.9 & 230.2 & 228.1 & 229.4\\
 & 32 & 197.5 & 222.8 & 232.1 & 251.2 & 247.2 & 251.1 & 248.6 & 247.5\\
 & 64 & 150.0 & 177.0 & 194.8 & 191.7 & 185.8 & 188.8 & 186.5 & 186.4\\
\midrule
\multirow{6}{*}[-0.5em]{Naive} & 0 & 154.5 & 154.3
	& 152.5 & 161.6 & 168.0 & 170.5 & 179.4 & 177.4\\
 & 4 & 182.5 & 180.0 & 178.0 & 189.8 & 193.4 & 198.3 & 202.2 & 200.7\\
 & 8 & 211.0 & 212.0 & 206.1 & 222.4 & 231.4 & 235.0 & 232.8 & 232.3\\
 & 16 & 256.0 & 253.8 & 257.3 & 270.5 & 276.3 & 287.1 & 282.8 & 280.9\\
 & 32 & 276.0 & 274.0 & 275.9 & 307.3 & 309.3 & 309.8 & 308.9 & 307.1\\
 & 64 & 183.0 & 205.0 & 235.4 & 234.7 & 228.7 & 229.2 & 227.4 & 229.2\\
\midrule
\multirow{6}{*}[-0.5em]{\fase} & 0 & 122.5 & 122.8 
	& 121.6 & 123.9 & 130.5 & 134.9 & 140.4 & 140.5\\
 & 4 & 161.0 & 160.3 & 155.0 & 163.7 & 167.9 & 172.4 & 175.4 & 172.5\\
 & 8 & 186.5 & 187.0 & 186.9 & 196.1 & 200.4 & 208.6 & 207.6 & 208.3\\
 & 16 & 212.5 & 218.0 & 216.1 & 234.2 & 240.1 & 241.4 & 242.0 & 241.7\\
 & 32 & 247.0 & 251.5 & 257.5 & 273.1 & 270.7 & 275.6 & 272.8 & 271.7\\
 & 64 & 171.0 & 195.3 & 223.8 & 214.7 & 210.3 & 212.2 & 211.6 & 210.2\\
\bottomrule
\end{tabular}
\vspace{-10pt}
\end{table}

\begin{figure}[tb]
\centering
\includegraphics[width=.6\columnwidth]{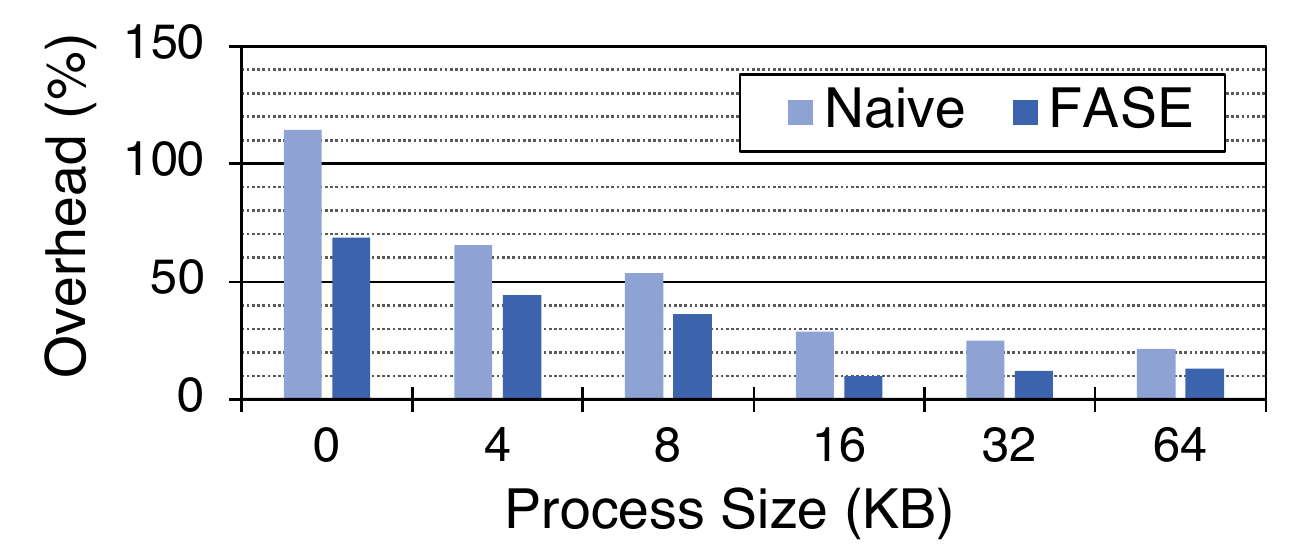}
\vspace{-14pt}
\caption{\latctx\ overhead vs. process sizes.}
\label{fig:latctxWss}
\vspace{-15pt}
\end{figure}

\subsection{Linux Kernel Results}\label{sec:krres}

Table~\ref{tab:latctx_full} shows the context switch latency of the baseline, naive, and \fase\ systems regarding different process sizes and numbers. Column 1 and Column 2 shows the system names and program sizes in KB. Columns 3 to 8 are the context switch latency in microseconds. Overall, smaller process sizes lead to lower context switch latency, since cache footprint is affected by the process size. The lowest context switch latency is witnessed when the baseline system is used with 0 KB process size, because it does not enforce flush of the cache and cache footprint is minimum. The highest context switch latency is found when the naive system is used with 32 KB process size, because full cache flush is enforced, and the cache footprint is maximum in this case. Process size of 64 KB is larger than cache size (32 KB) and shows smaller context switch time. This phenomenon is because the behavior of array accumulation in \latctx{} test and cache collision together lead to a ``friendly'' cache footprint at the context switch point.

Fig.~\ref{fig:latctxWss}  depicts the overhead of context switch latency as a function of process size. 
For each process size, the \latctx\ overhead is calculated from the geometric mean of the \latctx\ overhead among the process numbers. In general, when process size is equal to or smaller than cache size, the overhead decreases as process size increases. 
We also found process size affects context switch latency differently in \fase\ and naive systems. For \fase\ system, 16 KB process size results in the lowest overhead, followed by 32 KB and 64 KB. For the naive system, 64 KB process size has the lowest overhead, while 32 KB and 16 KB follow. This result is because \fase's LLSF mechanism can save half the cache from flushing if the content in the cache is equally owned by two processes. \fase\ shows the largest overhead reduction in percentage in comparison to the naive method when process size is 16 KB. 
On average, LLSF reduces context switch latency overhead of naive system by 42\%. In comparison to the naive method, LLSF reduces 66\% context switch latency overhead in the best case (when process size is 16 KB).
In summary, Linux kernel results demonstrate that \fase\ can significantly reduce mitigation overhead during OS execution.


\subsection{Microbenchmark Results}\label{sec:mbres}

\begin{figure}[tb]
\vspace{-8pt}
\centering
\includegraphics[width=.86\columnwidth]{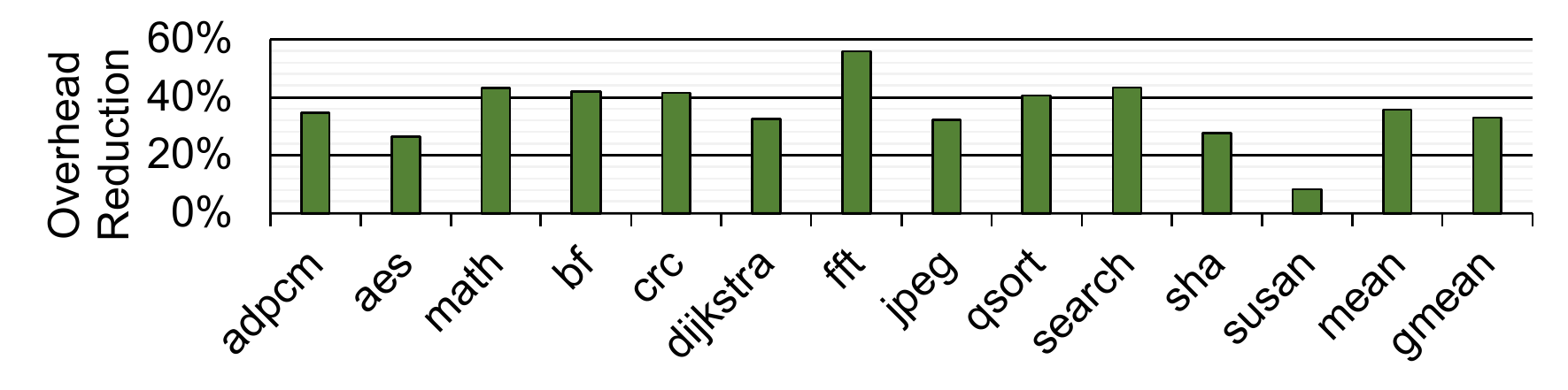}
\vspace{-14pt}
\caption{Execution time overhead reduction in percentage of \fase\ on top of naive system across MiBench programs (gmean: geometric mean).}
\label{fig:usrTime}
\vspace{-18pt}
\end{figure}

\begin{figure}[tb]
\centering
\subfloat[][Call graph \& CLSF scopes]{
\includegraphics[width=.36\columnwidth]{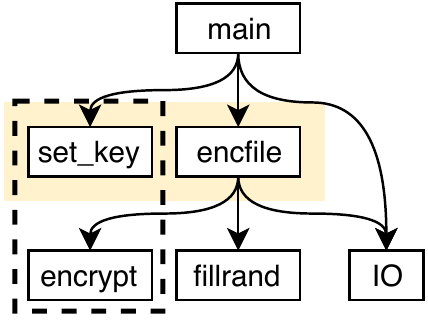} 
\label{fig:aesGraph}
}
\subfloat[][Execution time versus CLSF schemes]{
\includegraphics[width=.6\columnwidth]{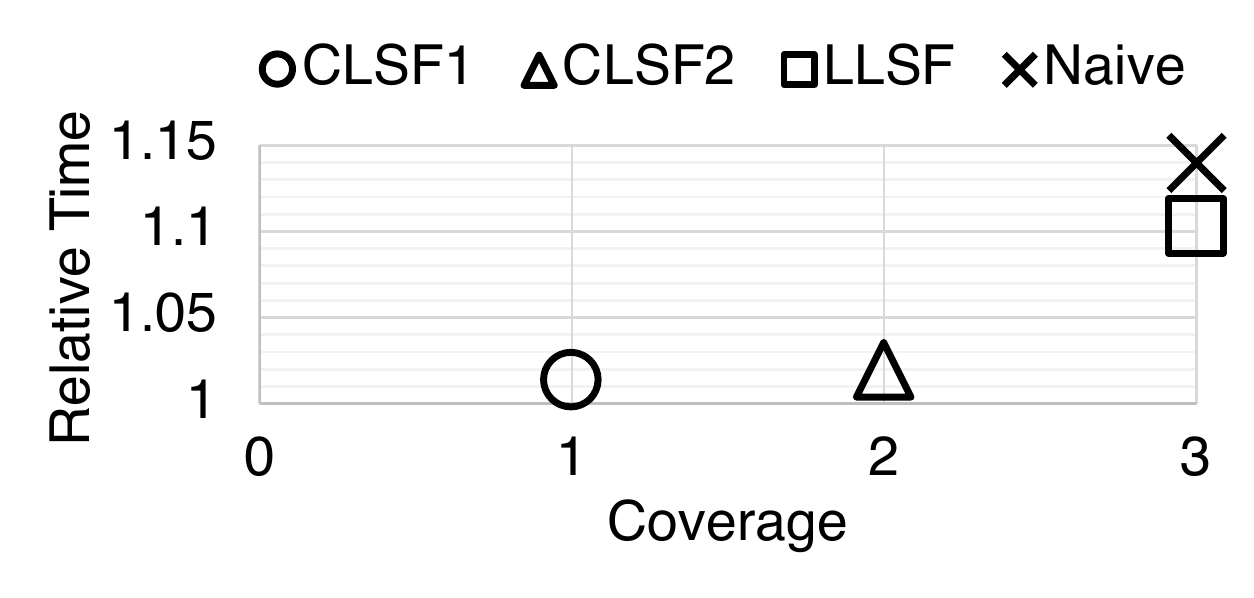} 
\label{fig:aesTime}
}
\vspace{-8pt}
\caption{\fase\ CLSF AES case study. Relative execution time is normalized on baseline AES. \reg{CLSF1}: \reg{set\_key} and \reg{encfile} in yellow background. \reg{CLSF2}: functions in dashed box in \protect\subref{fig:aesGraph}.}
\label{fig:fcfAes}
\vspace{-15pt}
\end{figure}

Fig.~\ref{fig:usrTime} shows how much overhead due to flush-based mitigation in the naive method is reduced by using \fase\ (LLSF alone). Overall, \fase\ reduces the execution time overhead effectively by 36\% on average (geometric mean is 33\%).
Since mibench programs mostly target embedded applications, some programs have few syscalls and/or a small cache footprint. Hence, the overhead in these programs can be very small, such as \reg{susan}. For other programs, \fase\ reduces the overhead significantly by around 30\% to 56\% from the naive method. 
\fase's average time overhead is 7.6\% (geometric mean is 3.4\%).

\textbf{CLSF AES case study.} AES (also called rijndael), which is a security-critical application, from MiBench, is used to observe the time saving of \fase's CLSF. Fig.~\ref{fig:fcfAes} presents the schemes and results of this case study. Fig.~\ref{fig:aesGraph} shows the call graph, which consists of the major functions in AES encryption. Based on this call graph, we created two choices of CLSF critical segments.
First, CLSF1 denotes CLSF covering \reg{set\_key} function and the main while-loop in \reg{encfile} function which calls \reg{encrypt}. This scope is denoted as ``1'' on the X-axis in Fig.~\ref{fig:fcfAes}. Second, CLSF2 covers \reg{set\_key} function and \reg{encfile} function. CLSF2 uses a larger scope as \reg{encfile} function is the parent function of \reg{encrypt}. This scope is denoted as ``2'' on the X-axis in Fig.~\ref{fig:fcfAes}. We compared CLSF1 and CLSF2 to LLSF and naive methods. The coverage of LLSF and naive methods is of the full program, denoted as ``3'' on the X-axis in Fig.~\ref{fig:fcfAes}.
Fig.~\ref{fig:aesTime} compares the relative execution time of these schemes. The relative execution time is calculated by normalizing the baseline system. 
It can be seen that \fase\ CLSF can reduce execution time greatly (to about 1\% overhead) from around 1.1 (equivalent to 10\% time overhead, LLSF) and 1.15 (equivalent to 15\% time overhead, naive). 
Among the CLSF schemes, CLSF2 takes slightly more time than CLSF1, due to larger critical segment coverage.


\subsection{Hardware Results}\label{sec:hwres}

\begin{table}\footnotesize
\centering
\caption{FPGA utilization across hierarchies.}\label{tab:fpgaUtil}
\vspace{-8pt}
\begin{tabular}{c c c c c c c }
\toprule
Subject & Resource & Tile & Core & DCache & Frontend & Tile[-FPU]  \\
\midrule 
\multirow{2}{*}{Baseline} & LUT & 30953 & 5451 & 2820 & 4844 & 14855 \\
						& F/F & 13916 & 2004 & 2596 & 4751 & 10125\\
\midrule
\multirow{2}{*}{\fase} & LUT & 30997 & 5457 & 2877 & 4861 & 14912 \\
					& F/F & 13991 & 2067 & 2607 & 4751 & 10198 \\
\midrule
\multirow{2}{*}{Overhead} & LUT & 0.1\% & 0.1\% & 2.0\% & 0.4\% & 0.4\% \\
						& F/F & 0.5\% & 3.1\% & 0.4\% & 0.0\% & 0.7\% \\
\bottomrule
\end{tabular}
\vspace{-16pt}
\end{table}



To understand the hardware cost of \fase, \fase\ processor is compared to the baseline processor. \fase\ processor has both LLSF and CLSF implemented. 
Overall, the only overhead of \fase\ in resource utilization is found in FPGA look-up tables (LUTs) and flip-flops (F/Fs). The block RAM utilization is unchanged, because \fase\ only adds a few hundred bits in the tag array in the cache. The max clock speed (clock frequency) is unchanged as the baseline processor, since \fase's hardware does not change the critical timing path. 

Table~\ref{tab:fpgaUtil} shows the FPGA resource utilization overhead of \fase. 
Since the entire SoC includes many large uncore components, a direct comparison at the SoC level will show negligible differences. To make a finer comparison, we focus on the overhead at hierarchies at and beneath the tile. Note that a  tile (\reg{RocketTile}) is a processor core and its local private resources (caches, TLBs, etc.).
In the tile, the resource utilization of the key components, core, L1 D-cache (\reg{DCache}), and frontend (I-cache and other instruction fetch components) are compared. Because FPU is quite a large component in tile (and remains unchanged), a comparison is made of the tile resource utilization with FPU's utilization removed (written \reg{Tile [-FPU]}).
As shown in Table~\ref{tab:fpgaUtil}, \fase\ increases 0.1\% LUTs and 0.5\% F/Fs. Without the FPU, this overhead becomes 0.4\% LUTs and 0.7\% F/Fs. \fase\ increases LUTs in the cache by 2\%. The 3\% F/F overhead in the core is caused by cross-boundary optimization of Vivado synthesis. These additional F/Fs are mostly from the \fase\ hardware in the cache.
 
 \begin{figure}[tb]
\centering
  \subfloat[b][Baseline]{
\includegraphics[width=.37\columnwidth,valign=b]{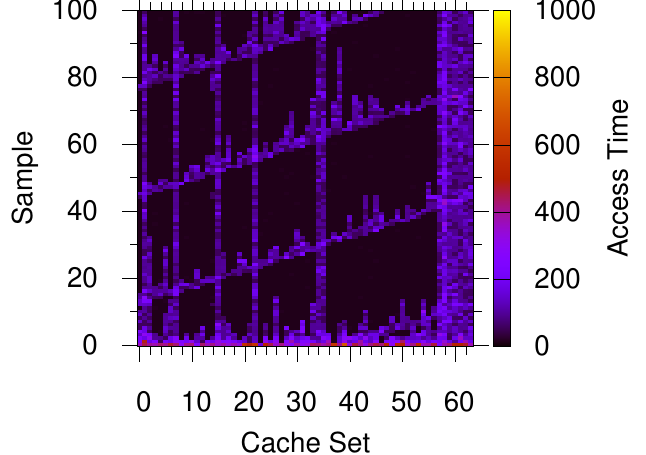} 
\label{fig:mastikBs}
  }
  \subfloat[b][Naive]{
  \includegraphics[width=.31\columnwidth,valign=b]{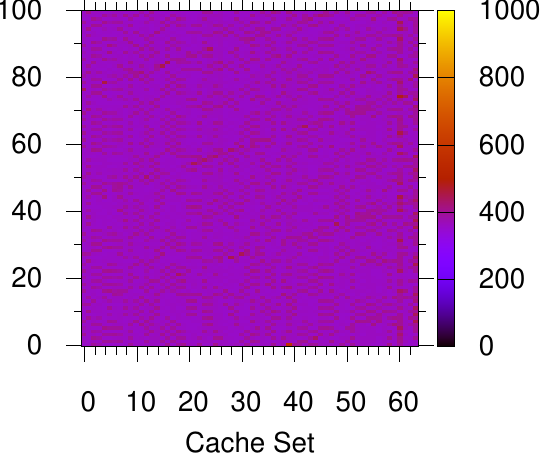} 
 \label{fig:mastikFx}
  }
  \subfloat[b][\fase]{
  \includegraphics[width=.31\columnwidth,valign=b]{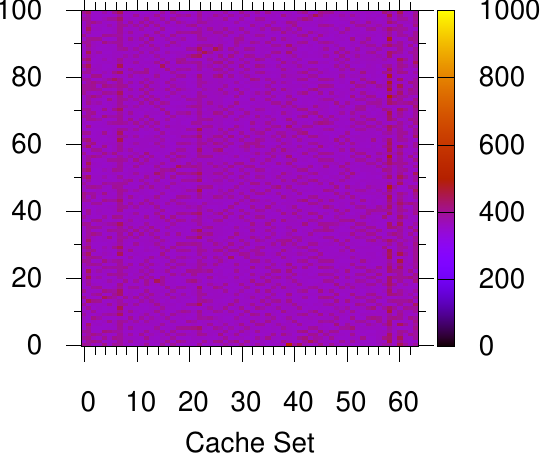} 
 \label{fig:mastikFase}
  }
  \vspace{-10pt}
\caption{\pp\ attack results on the baseline system, naive mitigation, and \fase\ mitigation. 
}
\label{fig:mastik}
\vspace{-16pt}
\end{figure}

\subsection{Security Results}\label{sec:secres}

Fig.~\ref{fig:mastik} uses heat maps to depict the \pp\ (a representative contention-based cache timing attack) results of the three systems (the baseline system without protection in Fig.~\ref{fig:mastikBs}, the naive mitigation system in Fig.~\ref{fig:mastikFx}, and the proposed \fase\ mitigation system~\ref{fig:mastikFase}). \fase\ LLSF and CLSF (covering victim function) showed equivalent effects in this experiment. In each figure, the X-axis is the cache set from 0 to 63. Y-axis is the sample number. One hundred samples are shown in these figures, which are sufficient to illustrate the afforded protection. The Z-axis (heat color) stands for the access time in clock cycles, taken for probing the cache sets. As a result,  without protection, the baseline system's victim cache access patterns can be seen (below 100 cycles shown in deep purple and black). As shown in Fig.~\ref{fig:mastikFx} and \ref{fig:mastikFase}, on naive and \fase\ systems, \pp\ observes misses in all cache sets (more than 100 clock cycles, similar colors across cache sets). This result shows that \fase's flush mechanism is effective for countering contention-based on-core cache timing attacks.

\section{Conclusion}\label{sec:eop}
In this paper, we have presented a novel flush-based method, \fase, to mitigate contention-based cache timing attacks. \fase\ leverages an ISA extension and modified cache microarchitecture to minimize flush cost of the mitigation. Our experiment shows that \fase\ can mitigate contention-based timing attacks with negligible hardware cost while reducing time overhead by 36\% for user programs and 42\% for OS context switch, in comparison to naive method.

\section*{Acknowledgement}
This research was supported by the Australian Research Council's Discovery Projects funding scheme (project DP190103916). We would like to thank Defence Science and Technology Group Australia for their support.

\printbibliography


\end{document}